\documentclass[doublecol]{epl2}

\usepackage[colorlinks,bookmarks=false,citecolor=blue,linkcolor=red,urlcolor=blue]{hyperref}

\title{Unconventional magnetization plateaus in a Shastry-Sutherland spin tube}
\author{S.R. Manmana\inst{1}\footnote{Present address: JILA (University of Colorado and NIST), and Department of Physics, University of Colorado at Boulder, CO 80309-0440, USA.} \and J.-D. Picon\inst{1} \and K.P. Schmidt\inst{2} and F. Mila\inst{1}}

\institute{
	\inst{1} Institut de Th\'eorie des Ph\'enom\`enes Physiques, EPF Lausanne, CH-1015 Lausanne, Switzerland \\
	\inst{2} Lehrstuhl f\"ur Theoretische Physik I, Technische Universit\"at Dortmund, D-44221 Dortmund, Germany, EU
}

\date{\today}


\abstract{
Using density matrix renormalization group (DMRG) and perturbative continuous unitary transformations (PCUTs),
we study the magnetization process in a magnetic field for all coupling strengths of a quasi-1D version of the 2D Shastry-Sutherland lattice, a frustrated spin tube
made of two orthogonal dimer chains. 
At small inter-dimer coupling, plateaus in the magnetization appear at 1/6, 1/4, 1/3, 3/8, and 1/2.
As in 2D, they correspond to a Wigner crystal of triplons. However, close to the boundary of the product singlet phase,
plateaus of a new type appear at 1/5 and 3/4. They are stabilized by the 
localization 
of {\it bound states} of triplons. Their magnetization profile differs significantly from that of
single triplon plateaus and leads to specific NMR signatures. 
We address the possibility to stabilize such plateaus in further geometries by analyzing small finite clusters using exact diagonalizations and the PCUTs.
}

\begin{document}

\maketitle
\section{Introduction}
\label{sec:introduction}

The competition between kinetic energy and inter-particle repulsion is a central theme of strongly correlated systems and leads to many interesting phenomena. 
In bosonic lattice models, the ground state can be insulating, superfluid or supersolid depending on the strength of the interaction and on the density \cite{murthy,QMC_triangular1,QMC_triangular2,QMC_triangular3}. 
At fractional filling, the insulating phases identified so far in lattice models are Wigner crystals in which elementary 
particles form a superlattice. 
A high-commensurability example has been observed a few years ago in a frustrated quantum magnet, SrCu$_2$(BO$_3$)$_2$\cite{miyahara03R}. 
\begin{figure}[t]
\includegraphics[width=0.49\textwidth]{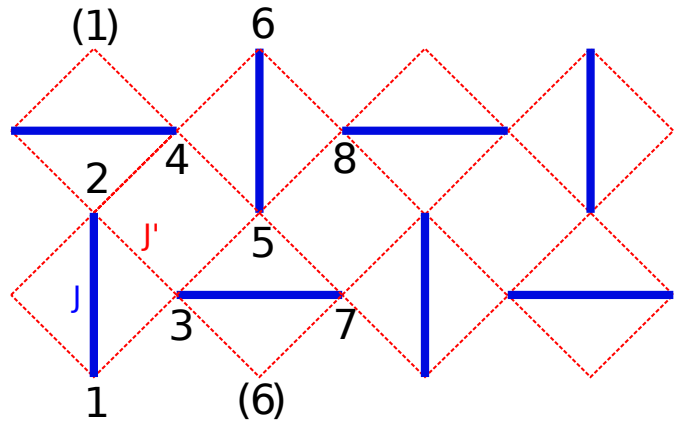}
\caption{(Color online) Graphical representation of the system.}
\label{fig:system}
\end{figure}
This dimer-based 2D antiferromagnet  is a realization of the Shastry-Sutherland lattice, and the bosonic particles are triplets injected by the magnetic field into the singlet product ground state.
They have a very small kinetic energy due to the frustrated nature of the inter-dimer interaction,  
and the 1/8 insulating state (identified in NMR experiments to be a crystal of triplet excitations \cite{kodama02}) is the first of a series of insulating phases, or magnetization plateaus in the magnetic language. 
The actual series is still controversial\cite{kageyama,sebastian}, but according to the latest NMR and magnetization results \cite{takigawa}, insulating phases occur at density 1/8, 2/15, 1/6, 1/4, and 1/3 (and probably 1/2\cite{sebastian}). 
Although no theory has been able so far to reproduce this series, the current belief is that all these phases are Wigner crystals of triplets. 
This is based on weak coupling approaches which have allowed one to reliably explore the parameter range $J'/J\leq 0.5$ \cite{dorier08,abendschein08} (where $J'$ and $J$ are the
inter- and intra-dimer couplings, see fig.~\ref{fig:system}), and to reproduce a series of magnetization plateaus similar (but not identical) to the experimental one.
On the other hand, it has been found that elementary excitations of the Shastry-Sutherland lattice can be bound states of triplet excitations \cite{knetter00_2a,knetter00_2b}, and the question arises, if in a strong magnetic field Mott insulators of such bound states can be stabilized. 
Since the reliability of the perturbative approaches is limited to $J'/J \le 0.5$, and in view of the significant
changes of the plateau structure upon increasing $J'/J$ reported in Ref.~\cite{dorier08}, 
investigations aimed at studying the magnetization for stronger inter-dimer couplings are clearly called for.
This is of particular relevance for SrCu$_2$(BO$_3$)$_2$, where $J'/J \approx 0.65$, quite close to the critical
ratio $J'/J\simeq 0.7$ where the product of singlets is no longer the ground state. \\

In this letter, we address these issues in the context of a quasi-1D version of the model,
a system of two coupled orthogonal dimer chains with periodic boundary conditions (PBC) applied in the transverse direction, shown in Fig.~\ref{fig:system}, which we call a {\it Shastry-Sutherland spin tube}.
On this geometry, we study the Heisenberg spin-$1/2$ Hamiltonian in an external magnetic field $H$,
\begin{equation}
\mathcal{H} = J \sum\limits_{\ll i,j \gg} \vec{S}_i \cdot \vec{S}_j  + J' \sum\limits_{<i,j>} \vec{S}_{i} \cdot \vec{S}_{j} - H \sum\limits_i S^z_i
\label{eq:ham}
\end{equation}
with the bonds $\ll i,j \gg$ building an array of orthogonal dimers and the bonds $<i,j>$ representing inter-dimer couplings. 
The main advantage over the 2D lattice is the possibility to use DMRG \cite{dmrg1,dmrg2}, which, as we shall see, provides very accurate results for all parameters, and has led to the identification of plateaus of a new type close to the boundary of the singlet phase, while the interpretation of the phase diagram, including the new plateaus, relies heavily on the PCUT approach \cite{knetter00a1,knetter00a2,knetter00_2a,knetter00_2b,dorier08}.

\section{Properties of the System and Methods}
\label{sec:methods}

Let us first review some simple consequences of the peculiar topology of the lattice. As for the 2D Shastry-Sutherland model, the product of singlets on all dimers is an eigenstate of the Hamiltonian\cite{miyahara03R}. But there are many more product
eigenstates, all consequences of the fact that the coupling of a vertical dimer to its neighbouring
horizontal dimers involves only its total spin. This is only true for two of them in 2D, but here it is
also true for the third one which lies above or below (see Fig.~\ref{fig:system}) because of periodic boundary
conditions. A vertical singlet is thus completely disconnected from the rest, and eigenstates can be
constructed as the product of an arbitrary number of vertical singlets times an eigenstate of the Hamiltonian
of the remaining sites. By the same token, a horizontal triplet surrounded by three vertical singlets can be factorized. By contrast, a vertical triplet can delocalize on its two horizontal neighbours on the same chain, and it is only
this three dimer unit that can factorize when surrounded by singlets. Finally, when $J=0$, the Hamiltonian can
be rewritten
\begin{equation}
\mathcal{H} = J' \sum\limits_{i \ {\rm odd}} (\vec{S}_i+\vec{S}_{i+1})\cdot (\vec{S}_{i+2}+\vec{S}_{i+3})- H \sum\limits_i S^z_i .
\label{eq:ham2}
\end{equation}
It depends only on the total spin $(\vec{S}_i+\vec{S}_{i+1})$ of pairs of sites with $i$ odd. These total spins are thus
conserved quantities and define sectors according to whether they are equal to 0 or 1. For all fields, the ground state is in the sector where they are equal to 1, as for the fully frustrated spin-1/2 ladder\cite{HoneckerMilaTroyer}, and the magnetization curve is exactly that of a Haldane chain \cite{haldane1,haldane2,kashurnikov99}.

\begin{figure}[t]
\includegraphics[width=0.49\textwidth]{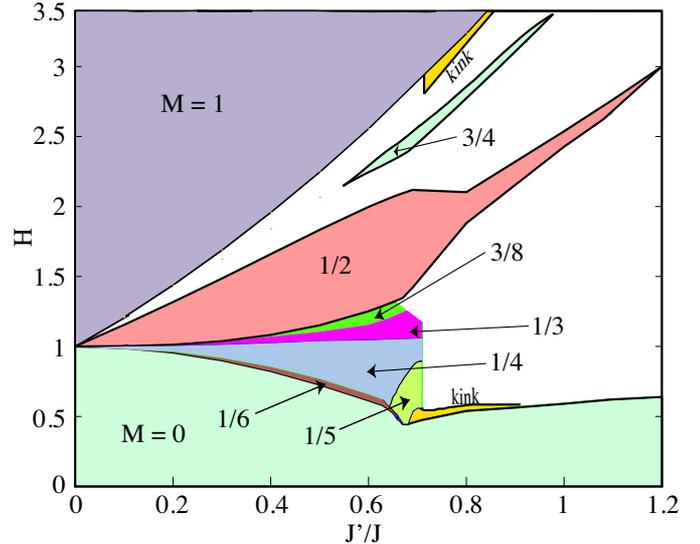}
\caption{(Color online) Sketch of the phase diagram of a Shastry-Sutherland tube in a magnetic field as obtained by our combined DMRG and PCUTs analysis. The various plateaus are described in more detail in the text.}
\label{fig:phasedia}
\end{figure}

Let us now give some comments on the methods. The effective hardcore boson model has been derived by PCUTs about the limit $J'/J=0$ as described in Ref.~\cite{dorier08}. According to the previous discussion, single triplets on horizontal dimers cannot move while single triplons on vertical dimers are localized on three-dimer units and
lower their energy by 
virtual fluctuations within this three-dimer unit. This results in a lower chemical potential for vertical dimers and the magnetization for $M<1/2$ is solely determined by these triplons (in the dimer phase). The effective model for vertical dimers contains only density-density interactions which have a finite range. Here we have  calculated all two-, three-, and four-triplon interactions. Note that kinetic terms are absent for vertical dimers and the Hartree approximation of the effective model becomes exact. By contrast, correlated hopping terms involving horizontal dimers are relevant for $M>1/2$.
For the DMRG, the lack of kinetic terms shows up as a tendency to get stuck in excited states.
In order to overcome this, additional fluctuations need to be included.
This can either be done ad-hoc in the course of the sweeps \cite{schmitteckert}, or by adding small anisotropic Dzyaloshinskii-Moriya (DM) interactions which break the SU(2) symmetry of the original problem.
We obtain a good agreement between both DMRG approaches and in the following we only quote the results obtained without DM anisotropies for systems with up to $L=360$ lattice sites.


\section{Phase diagram}
\label{sec:phasediagram}

The phase diagram as a function of field and inter-dimer coupling is depicted in Fig.~\ref{fig:phasedia}. It has been
obtained by combining PCUT and DMRG (see below). Three regimes of inter-dimer coupling can be identified. For small inter-dimer coupling, say up to $J'/J=0.5$, the magnetization consists of a series of plateaus at 1/6, 1/4, 1/3, 3/8 and 1/2. For large inter-dimer coupling ($J'/J>1.2$), the magnetization increases smoothly up to saturation. In between, there is an intermediate regime where a number of new features appear: kinks close to zero magnetization
and saturation\cite{kinks}, and more interestingly two new plateaus at 1/5 and 3/4.

\begin{figure}[t]
\includegraphics[width=0.49\textwidth]{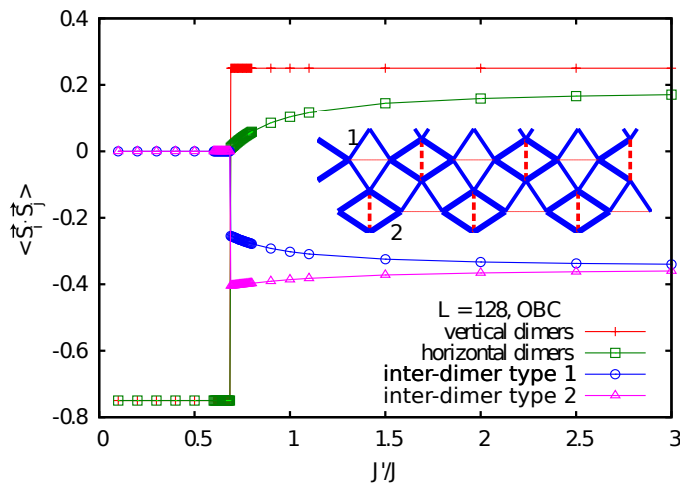}
\caption{(Color online) Spin-spin correlation $\langle \vec{S}_i \cdot \vec{S}_j\rangle$ as a function of $J'/J$ at zero magnetic field for the four different bonds in the bulk of the system. The inset shows the pattern on the bonds in the Haldane-like phase [blue (solid line): $\langle \vec{S}_i \cdot \vec{S}_j \rangle <0$, red (dashed line):  $\langle \vec{S}_i \cdot \vec{S}_j \rangle >0$].}
\label{fig:bonds}
\end{figure}

The change of behavior between small and large inter-dimer coupling can be
traced back to the zero-field phase diagram of the model. As for the 2D Shastry-Sutherland model,
for small enough $J'/J$, the product of singlets is the ground state. Upon increasing $J'/J$, there has to
be at least one phase transition since the low-energy sector of the $J=0$ model is equivalent to a Haldane chain.
It turns out that there is a single, first-order transition between two gapped phases, as clearly revealed
by DMRG calculations: spin-spin correlations have a dramatic jump at $J'/J\simeq 0.66$ (see Fig.~\ref{fig:bonds}), the ground state energy per site has a kink, the singlet-singlet gap closes at the same value, and the singlet-triplet gap never closes (results not shown),
while beyond that value the correlations smoothly evolve towards the values of a spin-1 chain.
So the intermediate regime cannot be simply explained by a different zero field ground state
in this parameter range.

\begin{figure}[t]
\includegraphics[width=0.49\textwidth]{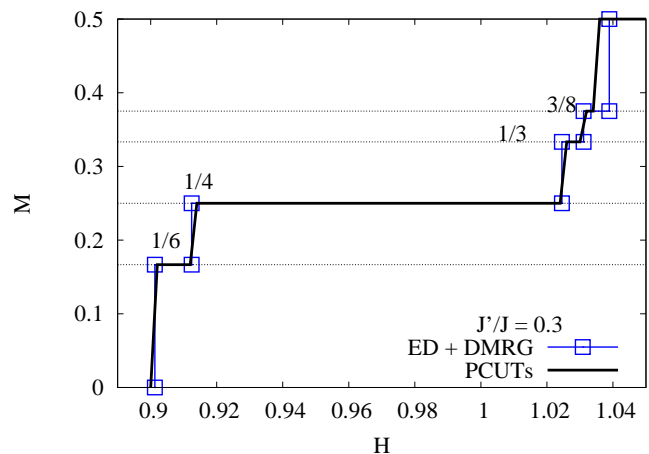}
\caption{(Color online) Magnetization curve for $J'/J=0.3$ below $M=1/2$ as obtained by exact diagonalizations and DMRG calculations (s. text) and by PCUTs+Hartree approximation.}
\label{fig:magnetizations0.3}
\end{figure}

\subsection{Small inter dimer coupling}
\label{sec:smallJp}
Let us now have a closer look at the plateaus in the small $J'/J$ regime. In the domain of applicability of PCUT
(not too large $J'/J$, density not larger than 1/2), the agreement between PCUT and DMRG is truly excellent,
as can be seen on Fig.~\ref{fig:magnetizations0.3}, where PCUT and DMRG results are compared for $J'/J=0.3$.
The only noticeable discrepancy concerns the width of the 3/8 plateau. 
We believe that this might be due to the presence of an infinite sequence of plateaus, as in the orthogonal dimer chain \cite{koga00,schulenburg02}, that cannot be captured by our techniques. Indeed, vertical triplons have no kinetic energy, as in the 1D case. At the same time, PCUT shows that high order corrections lead to repulsions of increasing range, suggesting that the 'exact' effective model has long-range repulsions. Together with the rigorous absence of kinetic energy, this is expected to lead to a sequence of plateaus of increasing commensurability. Note, however, that this sequence cannot be reproduced quantitatively neither by PCUT, which cannot calculate the repulsion beyond a certain range, nor by DMRG, which is limited to finite sizes. 

\begin{figure}[t]
\includegraphics[width=0.49\textwidth]{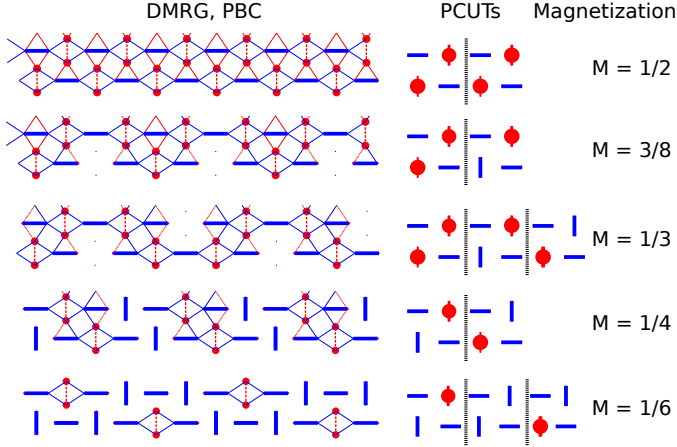}
\caption{(Color online) Structure of the magnetization plateaus resp. Mott insulating phases shown in Fig.~\ref{fig:magnetizations0.3} obtained by the DMRG (left side) resp. by the PCUTs for $J'/J=0.3$, representative for the parameter range $J'/J < 0.66$ and $M \leq 1/2$.  The colors red (dashed bonds) and blue (solid bonds) indicate a positive or negative value of the spin-spin correlation $\langle \vec{S}_i \cdot \vec{S}_j\rangle$ and of the local magnetizations $\langle S_i^z\rangle$, respectively. The circles in the PCUTs column indicate the dimers occupied by a triplon.}
\label{fig:plateauxstructures}
\end{figure}

The structures of these plateaus are also in perfect agreement (see Fig.~\ref{fig:plateauxstructures}). In the
bosonic (resp. magnetic) language, they consist of Mott insulating phases with well localized bosons (resp. triplons), in qualitative agreement
with phases detected so far in the 2D Shastry-Sutherland model. Note that all plateaus correspond
to product wave-functions with vertical dimer singlets. This has been first observed by DMRG calculations including
all sites, then used systematically to improve the accuracy of DMRG by performing calculations with those
dimers removed. For the 1/6 and 1/4 plateaus, these singlets actually cut the tube into independent finite
segments, and the energy could be calculated exactly by diagonalizing very small clusters.

\subsection{Intermediate couplings: Wigner crystal of bound triplons}
\label{sec:intermediateJp}
\begin{figure}[t]
\includegraphics[width=0.49\textwidth]{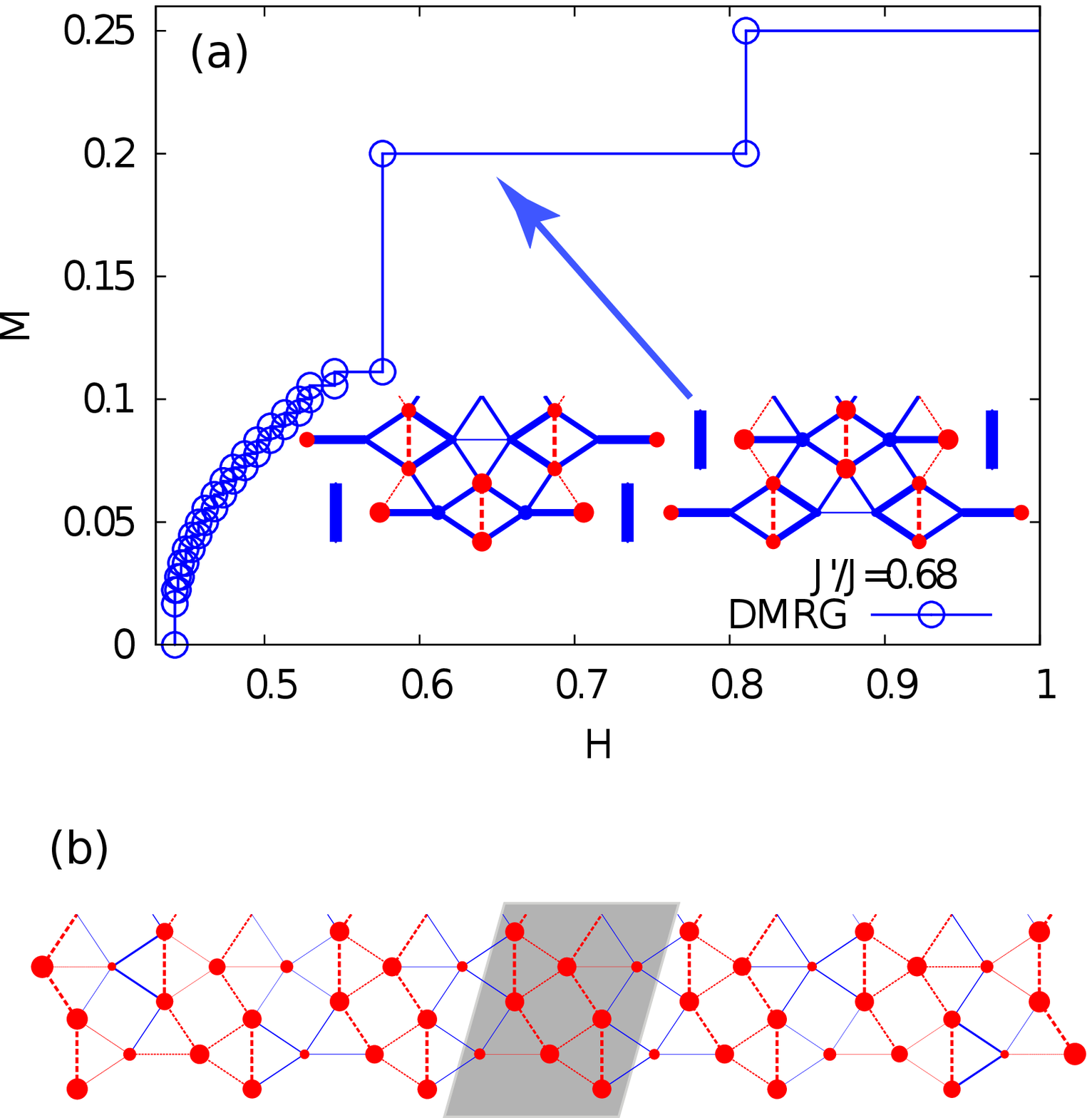}
\caption{(Color online) (a) Magnetization curve at $J'/J = 0.68$ at low fields as obtained by the DMRG. The inset shows the structure of the 1/5-plateau. (b) Structure of the magnetization plateau at 3/4 for $J'/J = 0.68$ as obtained by the DMRG. The
shaded area highlights the four-dimer unit cell.}
\label{fig:15plateau}
\end{figure}

By contrast, the plateaus that appear in the intermediate regime are of a different nature. Let us concentrate on the 1/5 plateau. The magnetization curve at $J'/J=0.68$ is depicted in Fig.~\ref{fig:15plateau}(a), and the structure
of this plateau is shown as an inset. This plateau is again a product state of finite segments, so its energy
and internal structure are known exactly from diagonalizing a 16-site cluster. What is remarkable is
that it does not correspond to localized triplons. Indeed, the magnetization is spread over {\it three}
vertical dimers, although the total $S^z$ is equal to 2. This should be contrasted to the plateaus
at small inter-dimer coupling, which all correspond to localized triplons on vertical dimers.
This difference can be made more precise with the help of the PCUT analysis, which by essence
is able to keep track of the number of triplons in any eigenstate. For the 16 site building
brick of the 1/5 plateau, the ground state inside the 1/5 plateau has $S_{\rm total} = S^z_{\rm total} = 2$.
The PCUT analysis reveals that there are strong attractive interactions, and that this eigenstate is adiabatically connected to an excited three-triplon state of the lattice with $S=2$ in the limit $J'=0$, as shown in fig.~\ref{fig:traceback_state}.
This is the main finding of the present letter: in the intermediate region $0.65 \stackrel{<}{\sim} J'/J \stackrel{<}{\sim} 0.7$, a product state of singlets and of an extended bound state of triplet excitations forms an exotic Mott-insulator at $M=1/5$. 
The translational symmetry inside the plateau is not broken by the freezing of localized single-triplon objects as previously identified in the dimer phase of the 2D Shastry-Sutherland lattice, but by the crystallization of extended 
triplon bound states which carry $S=2$. 
Below, we will analyze this bound state in more detail and discuss the possibility to observe such structures in the 2D system by analyzing various finite size clusters of different shapes.  

The 3/4 plateau also seems to be of a similar nature.
The unit cell contains 8 sites (4 dimers), and the magnetization $S^z = 3$ of a unit cell is spread over the 4 dimers
[see fig.~\ref{fig:15plateau}(b)] with no indication of an empty dimer but
rather of two half-polarized dimers. 
This suggests that this plateau might correspond to the condensation of
an $S = 3$ bound state of four triplons. However, the delocalization
of one of the triplons might also be due to correlated
hopping, which is present for magnetization larger than 1/2.
Which of the two possibilities is realized could not be
checked along the same lines as for the 1/5 plateau
since the wave function of the 3/4 plateau is not
a product state.

\begin{figure}
\includegraphics[width=0.49\textwidth]{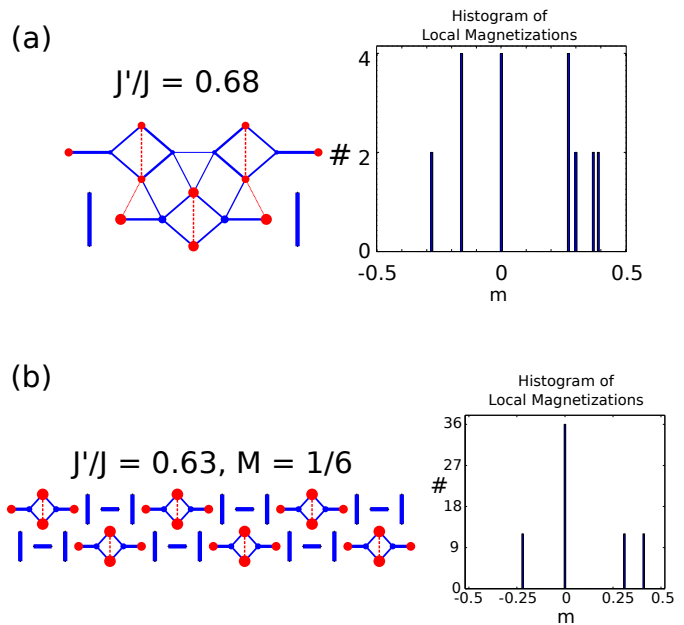}
\caption{(Color online) (a) Magnetization pattern and histogram of the magnetizations for the building block of the $M=1/5$ plateau at $J'/J=0.68$. 
(b) Typical magnetization pattern and histogram of magnetizations for a plateau which is a product state of singlets and triplons (here, the M=1/6 plateau at J'/J = 0.63). The results in (a) are obtained by ED, the ones in (b) by DMRG.  
}
\label{fig:histogram_butterfly}
\end{figure}

The magnetization pattern of these plateaus is very different from that of plateaus which are product states of single triplons and singlets.  
If $J'/J$ is large enough, a single localized triplon leads to a very characteristic magnetization pattern
on a 3-dimer unit already observed by NMR in the 1/8 plateau of SrCu$_2$(BO$_3$)$_2$\cite{kodama02}: the
magnetization is large, along the field, and of comparable magnitude on the sites of the central dimer and on the sites further apart, while it is opposite to the field on the first neighbors of the central dimer. 
In fig.~\ref{fig:histogram_butterfly}(b), the histogram of magnetizations of the plateau at $M=1/6$ is shown which is of this type.  
By contrast, a bound state of triplons leads to sites with intermediate polarizations along the field. 
This is for instance true for two of the three dimers of the building brick of the 1/5 plateau, whose polarization
is about twice as small as that of the third one.
The resulting histogram of magnetizations is shown in fig.~\ref{fig:histogram_butterfly}(a).  
In addition, one observes that the histograms of plateaus of single triplons have much more pronounced peaks than the ones of the bound states, which are flatter. This is due to the wider extension of the bound state on the cluster: for plateaus of single triplons at density $1/p$, 3 dimers out of $p$ are polarized in each unit cell since each triplon extends also over the two neighboring dimers. Hence, the density of unpolarized singlets is $(p-3)/p$, leading to a very strong peak in the histograms at $M=0$. This peak is significantly reduced in the case of bound states. As can be seen in fig.~\ref{fig:histogram_butterfly}, only two singlets remain unpolarized in the building block of the 1/5 plateau, while in the case of single triplons it would be four unpolarized singlets.  

\subsection{Analysis of the bound states}
\label{sec:analysisboundstates}

\begin{figure}[t]
\includegraphics[width=0.485\textwidth]{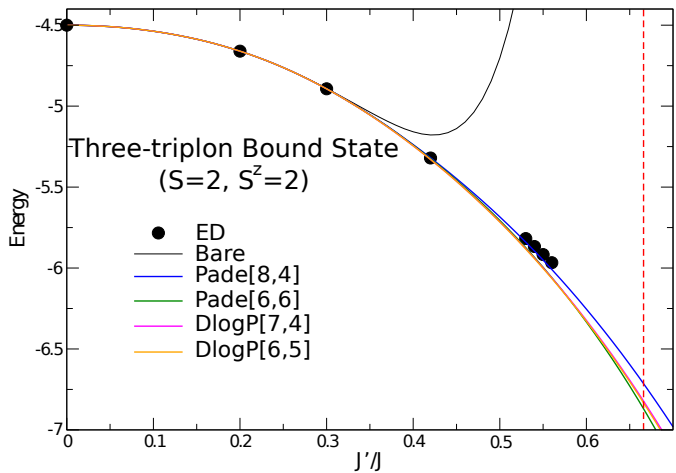}
\caption{(Color online) Energies for the bound state of fig.~\ref{fig:15plateau}(a) as obtained by ED and by the PCUTs, showing that the structure is adiabatically connected to a three-triplet excitation at $J'/J=0$.}
\label{fig:traceback_state}
\end{figure}

\begin{figure}[t]
\includegraphics[width=0.485\textwidth]{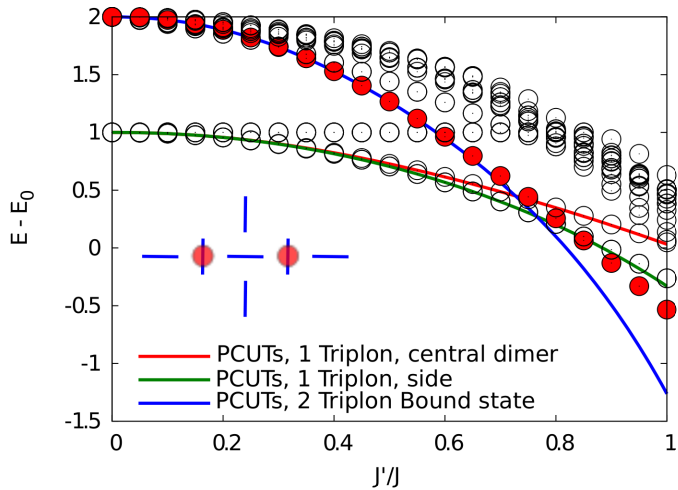}
\caption{(Color online) Energy levels in the sector $S^z_{\rm total} = 1$ of a small planar cluster of 7 dimers as  obtained by ED and by PCUTs (the inset shows the structure of the cluster; the red (filled) circles indicate the dimers occupied by the bound state). The circles represent the energies as obtained by ED, the red (filled) circles indicate the energies of the ED adiabatically connected to the 2-triplon $S=1$ bound state obtained by the PCUTs. 
}
\label{fig:smallclusters}
\end{figure}

As discussed above and shown in fig.~\ref{fig:traceback_state}, the building block of the $M=1/5$ plateau is adiabatically connected to a three triplet excitation at $J'/J=0$. 
This insinuates that the structure is a bound state of three triplons which, however, has to be contrasted to the value $S=2$ of the total spin carried by this structure. 
This value can be obtained by either coupling two $S=1$ objects to carry $S=2$ and then adding an additional triplet (leading to a substructure carrying $S=2$), or by adding two triplets to an object with $S=1$ and then coupling to an additional triplet. 
By computing the expectation value of $S^2$ on the two polarized vertical dimers on the same line and at the third polarized dimer independently we find $S=1$ for the two dimers on the same chain, and $S=1$ for the third one. 
This shows that the two vertical dimers on the same chain are in a superposition of a $|T_1\rangle$ and a $|T_0\rangle$ triplet state, which is then coupled to an additional $|T_1\rangle$ triplet state.  
The nature of the building block of the $M=1/5$ plateau is hence rather peculiar, and the question arises to which extend such a structure can be stabilized in a 2D system. 
In order to address this question, we have analyzed small 
clusters with open boundary conditions, two symmetric ones 
with 7 dimers (shown in fig.~\ref{fig:smallclusters}) and 11 dimers, and an asymmetric 
configuration of 13 dimers, by using exact diagonalizations and the 
PCUTs. It turns out that, although a bound state of three triplons is 
never stabilized in these geometries up to $J'/J = 0.7$, the
energy of a $S = 1$ bound state of a $|T_0\rangle$ and a $|T_1\rangle$ triplet
becomes comparable to that of the lowest single triplon states
around $J'/J=0.7$ for all clusters investigated. For the 7 dimer
cluster, this bound state even becomes the ground state for
$J'/J>0.8$. This suggests that plateaus of such two-triplon 
bound states are also viable candidates in the 2D lattice
for large enough $J'/J$. 

\section{Conclusions}
\label{sec:conclusions}
In conclusion, we have determined the phase diagram of a Shastry-Sutherland spin tube in a magnetic field using a combined PCUTs and DMRG approach.
In the limit of small $J'/J$ we find a sequence of fractional magnetization plateaus starting with 1/6, 1/4, 1/3 and 3/8, and a pronounced plateau at 1/2.
Due to the lack of kinetic terms we expect an infinite series of plateaus to be realized between 3/8 and 1/2.
In the limit of large $J'/J$, the system behaves as a spin-1 chain.
In the intermediate region, however, we find that the plateau at 1/6 is replaced by one at 1/5 with a complex structure {\it not} formed by local triplon excitations, but by extended $S=2$ triplon bound states, and that another plateau appears at 3/4 whose magnetization pattern is also incompatible with local triplons. 
The mechanism that leads to the stabilization of these plateaus of a new type
is not specific to 1D. 
It relies on the stabilization of a bound state of triplons, and bound states of two triplons
have been identified in the 2D system SrCu$_2$(BO$_3$)$_2$ \cite{miyahara03R} and in the Shastry-Sutherland model \cite{knetter00_2a,knetter00_2b}. 
The main specificity of the present calculation as compared to previous investigations in 2D is that we could access the parameter range close to the
boundary of the product singlet ground state. 
While the peculiar structure of three bound triplons remains specific for the Shastry-Sutherland tube, results for small clusters indicate that it is possible to realize 
competitive states 
of two bound triplons in the corresponding sector of $S^z_{\rm total}$. 
So we believe that such plateaus could appear in the Shastry-Sutherland and further 2D dimer based frustrated quantum magnets.
This is left for future investigation.

\acknowledgements
We acknowledge enlightening discussions about plateaus in SrCu$_2$(BO$_3$)$_2$ with C. Berthier, M. Horvati\'c and M. Takigawa, and J. Dorier for help with the numerical implementation of the Hartree approximation. 
SRM acknowledges financial support by PIF-NSF (grant No. 0904017), and 
KPS acknowledges ESF and EuroHorcs for funding through his EURYI. This work has been supported by the SNF and by MaNEP.


\begin{thebibliography}{10}
\expandafter\ifx\csname url\endcsname\relax\def\url#1{\texttt{#1}}\fi

\bibitem{murthy}
\Name{Murthy G., Arovas D. \and Auerbach A.} \REVIEW{Phys. Rev. B
  }{55}{1997}{3104}.

\bibitem{QMC_triangular1}
\Name{Wessel S. \and Troyer M.} \REVIEW{Phys. Rev. Lett. }{95}{2005}{127205}.

\bibitem{QMC_triangular2}
\Name{Heidarian D. \and Damle K.} \REVIEW{Phys. Rev. Lett. }{95}{2005}{127206}.

\bibitem{QMC_triangular3}
\Name{Melko R.~G., Paramekanti A., Burkov A.~A., Vishwanath A., Sheng D.~N.
  \and Balents L.} \REVIEW{Phys. Rev. Lett. }{95}{2005}{127207}.

\bibitem{miyahara03R}
For a review, see \Name{Miyahara S. \and Ueda K.} \REVIEW{J. Phys.: Condens. Matter
  }{15}{2003}{R327} and references therein.

\bibitem{kodama02}
\Name{Kodama K., Takigawa M., Horvatic M., Berthier C., Kageyama H., Ueda Y.,
  Miyahara S., Becca F. \and Mila F.} \REVIEW{Science }{298}{2002}{395}.

\bibitem{kageyama}
\Name{Kageyama H., Yoshimura K., Stern R., Mushnikov N.~V., Onizuka K., Kato
  M., Kosuge K., Slichter C.~P., Goto T. \and Ueda Y.} \REVIEW{Phys. Rev. Lett.
  }{82}{1999}{3168}.

\bibitem{sebastian}
\Name{Sebastian S.~E., Harrison N., Sengupta P., Batista C., Francoual S., Palm
  E., Murphy T., Marcano N., Dabkowska H. \and Gaulin B.} \REVIEW{Proc. Natl. 
  Acad. Sci. }{105}{2008}{20157}.

\bibitem{takigawa}
\Name{Takigawa M. et al.} private communication.

\bibitem{dorier08}
\Name{Dorier J., Schmidt K.~P. \and Mila F.} \REVIEW{Phys. Rev. Lett.
  }{101}{2008}{250402}.

\bibitem{abendschein08}
\Name{Abendschein A. \and Capponi S.} \REVIEW{Phys. Rev. Lett.
  }{101}{2008}{227201}.

\bibitem{knetter00_2a}
\Name{Knetter C., B\"uhler A., M\"uller-Hartmann E. \and Uhrig G.~S.}
  \REVIEW{Phys. Rev. Lett. }{85}{2000}{3958}.

\bibitem{knetter00_2b}
\Name{Knetter C. \and Uhrig G.~S.} \REVIEW{Phys. Rev. Lett.
  }{92}{2004}{027204}.

\bibitem{dmrg1}
\Name{White S.~R.} \REVIEW{Phys. Rev. Lett. }{69}{1992}{2863}.

\bibitem{dmrg2}
\Name{Schollw\"ock U.} \REVIEW{Rev. Mod. Phys. }{77}{2005}{259}.

\bibitem{knetter00a1}
\Name{Knetter C. \and Uhrig G.~S.} \REVIEW{Eur. Phys. J. B }{13}{2000}{209}.

\bibitem{knetter00a2}
\Name{Knetter C., Schmidt K.~P. \and Uhrig G.~S.} \REVIEW{J. Phys. A: Math. 
  Gen. }{36}{2003}{7889}.

\bibitem{HoneckerMilaTroyer}
\Name{Honecker A., Mila F. \and Troyer M.} \REVIEW{Eur. Phys. J. B
  }{15}{2000}{227}.

\bibitem{haldane1}
\Name{Haldane F.} \REVIEW{Phys. Rev. Lett. }{50}{1983}{1153}.

\bibitem{haldane2}
\Name{Haldane F.} \REVIEW{Phys. Lett. A }{93}{1983}{464}.

\bibitem{kashurnikov99}
\Name{Kashurnikov V.A., Prokof'ev N.V., Svistunov B.V. \and Troyer, M.} \REVIEW{Phys. Rev. B}{59}{1999}{1162}. 

\bibitem{schmitteckert}
\Name{Schmitteckert P. \and Eckern U.} \REVIEW{Phys. Rev. B }{53}{1996}{15397}
  and references therein.

\bibitem{kinks}
These kinks will not be discussed further since they are
  not specific to this model. See e.g. \Name{Okunishi K., Hieida Y. \and Akutsu Y.} \REVIEW{Phys. Rev. B
  }{60}{1999}{R6953}.

\bibitem{koga00}
\Name{Koga A., Okunishi K. \and Kawakami N.} \REVIEW{Phys. Rev. B
  }{62}{2000}{5558}.

\bibitem{schulenburg02}
\Name{Schulenburg J. \and Richter J.} \REVIEW{Phys. Rev. B }{65}{2002}{054420}.

\end{thebibliography}
\end{document}